\documentclass[pra,showpacs,twocolumn]{revtex4-1}
\usepackage{bm,graphicx,amsmath, color}
\newcommand{\ahat}{\hat{a}}
\newcommand{\Ahat}{\hat{A}}
\newcommand{\psihat}{\hat{\psi}}

\begin{document}
\title{Delocalization and superfluidity of ultracold bosonic atoms in
  a ring lattice} 
\author{Fernanda Pinheiro$^{1,2,3}$}
\email{fep@fysik.su.se}
\author{A. F. R. de Toledo Piza$^{1}$}
\affiliation{$^1$Institute of Physics, University of S\~{a}o Paulo,
  S\~{a}o Paulo, Brazil}
\affiliation{$^2$Department of Physics, Stockholm University, 
SE-106 91 Stockholm, Sweden}
\affiliation{$^3$NORDITA, KTH Royal Institute of Technology and
  Stockholm University, Se-106 91 Stockholm, Sweden} 
\date{\today}

\begin{abstract}
  Properties of bosonic atoms in small systems with a periodic quasi
  one-dimensional circular toroidal lattice potential subjected to
  rotation are examined by performing exact diagonalization in a
  truncated many body space. The expansion of the many-body
  Hamiltonian is considered in terms of the first band Bloch
  functions, and no assumption regarding restriction to 
  nearest-neighbor hopping (tight-binding approximation) is
  involved. A finite size version of the zero temperature phase
  diagrams of Fisher et al.~\cite{Fisher} is obtained and the results,
  in remarkable quantitative correspondence with the results available
  for larger systems, discussed.  Ground state properties relating
  to superfluidity are examined in the context of two-fluid
  phenomenology. The basic tool, consisting of the intrinsic inertia
  associated with small rotation angular velocities in the lab frame,
  is used to obtain ground state `superfluid fractions'
  numerically. They are analytically associated with one-body, uniform
  solenoidal currents in the case of the adopted geometry. These
  currents are in general incoherent superpositions of contributions
  from each eigenstate of the associated reduced one-body densities,
  with the corresponding occupation numbers as weights. Full coherence
  occurs therefore only when only one eigenstate is occupied by all
  bosons. The obtained numerical values for the superfluid fractions
  remain small throughout the parameter region corresponding to the
  `Mott insulator to superfluid' transition, and saturate at unity
  only as the lattice is completely smoothed out.
\end{abstract}

\maketitle

\section{Introduction}
Following the amazing development of experimental techniques in the  
latest years, cold atoms systems became the
primary candidates for the study of many-body quantum
phenomena. Mean-field aspects of condensation, for example, have been
extensively investigated both at the experimental~\cite{Ketterle, Cornell}
and theoretical~\cite{Griffin, LiebBook, PethickBook, PitaevskiiBook,
  PizaLectures, LeggettBook} levels. Properties of the strongly
correlated regime became accessible with the use of optical
lattices, and the transition from Mott-insulator to superfluid
was verified in the lab~\cite{Greiner}: in the superfluid phase, the atoms
are delocalized in the lattice in a state with long-range coherence,
whereas in the insulator phase they are localized in the lattice
sites, each of these, with a fixed number of atoms~\cite{Greiner}. 

Fundamental aspects of superfluid behavior relating to flux properties
in systems of alkali gases are still a matter of active research. The
rigorous theoretical proof of superfluidity in the Gross-Pitaevskii
limit was established only in 2002~\cite{Lieb}, in terms of a criteria
based on inertia and two-fluid arguments. In the current
experimental scenario, different setups which include effects of
rotation of containers have been proposed~\cite{Osterloh, Cooper}
and/or realized~\cite{Perrin}. In particular, persistent flow in a
toroidal trap~\cite{Phillips} and frictionless flow on a 2D system
subjected to a moving obstacle~\cite{Dalibard} were observed very
recently. This highlights the importance of explicitly considering the
coupling of the system with its moving boundaries. In fact, in the
context of optical lattices, earlier works~\cite{Keith1, Keith2} have
already investigated superfluid properties in terms of the response of
the system to imposed phase twists within the Bose-Hubbard model. In
these studies, however, ``the influence of the lattice potential itself
on the superfluid flow'' has been neglected, and it is not clear how
inclusion of such influence affects the superfluid properties and
relates to the transition from localized to delocalized bosons in the
many-body ground state.

Motivated by this, we investigate here superfluidity properties in a
weakly interacting Bose gas trapped in a rotating annular toroidal
Kronig-Penney trap, which constitutes a simplification of the
experimentally implemented optical lattices. In order to study
superfluidity in terms of an inertial criterion, we consider an
externally imposed rotation (`cranking') of the lattice. The result is
a `cranked' extension of the field Hamiltonian underlying the 
Bose-Hubbard model. This Hamiltonian is conveniently expanded in a
single-particle basis of Bloch functions and single-particle energies,
and all the required two-body matrix elements are explicit calculated.
In the calculations reported below, the single-particle 
basis is truncated to the first band only. However, within the bounds set by
this limitation, this treatment effectively relaxes the
constraints inherent to the tight-binding approximation which
underlies the standard usage of the Bose-Hubbard model: the range of
the tunneling is not restricted to nearest neighbors, and
interactions are no longer restricted to occur onsite. It therefore
includes nontrivial many-body effects stemming from the multi-mode
treatment as e.g. cross-collisional induced effects~\cite{Campinas}.

In terms of the results of the numerical many-body diagonalization one
can then obtain superfluid fractions (in the sense of the two-fluid
model) and associated currents. These results are compared with
results obtained for a finite version of the zero temperature phase
diagram~\cite{Fisher} of the Mott-insulator to superfluid transition.
And despite being a small system, to which the strict definition of
phase transition does not apply, it still deploys `precursor' features
which can be related even quantitatively to those which have been both
observed in real systems and supported by approximate computational
results obtained for considerably larger systems. In this context,
properties relating to condensation are also analyzed. We apply the
Penrose-Onsager~\cite{Onsager} criteria, which considers the
establishment of occupation dominance of one of the eigenstates of the
reduced density matrix. In particular, it is seen that the closure of
the Mott lobes happens for the same range of parameter values at which
occupation dominance essentially attains its peak value, and that
these changes evolve in a scale different from that associated with
the overall changes of the superfluid fraction and superfluid current.

The paper is organized as follows: in Sec.~\ref{sec:sec2} we present
the cranked version of the lattice model, whose ground-state
properties are analyzed in Sec.~\ref{sec:phasediagram} in terms of the
zero-temperature phase diagrams. In sections~\ref{sec:superfluidity}
and~\ref{sec:condensation} we discuss properties related,
respectively, with superfluidity and condensation, and in
Sec.~\ref{sec:conc} we present our conclusions.

\section{The cranked quasi-momentum Hamiltonian}\label{sec:sec2}

\subsection{Bose-Hubbard model}

The many-body dynamics of cold bosonic atoms in external lattice
potentials is strongly dominated by the two basic ingredients
consisting of hopping and short range (repulsive) two-body interaction
effects~\cite{Jaksch}. In a tight-binding regime hopping is dominated
by nearest 
neighbor processes and two-body effects are dominated by `on-site'
contributions only. These ingredients are combined in the Bose-Hubbard
model Hamiltonian
\begin{eqnarray}\label{1}
H_{BH}&=& -J \sum_{\langle i,j\rangle}(\ahat_i^{\dagger}\ahat_j +
\ahat_j^{\dagger}\ahat_i) +\nonumber \\ && +\frac{U}{2}\sum_i 
\ahat^{\dagger}_i\ahat^{\dagger}_i\ahat_i\ahat_i, \hspace{0.5cm}
J,U\;>\; 0,
\end{eqnarray}
where the first sum is restricted to nearest neighbors $\langle
i,j\rangle$ and the 
operator $\ahat_i$ ($\ahat_i^{\dagger}$) destroys (creates) a bosonic
particle in site $i$. In order to connect this Hamiltonian involving
`sites' to a more basic description involving spatial coordinates, one
associates the sites to amplitudes defined in terms of the first band
Wannier function, which is sufficiently well localized in space in the
tight-binding regime. As is now well known, competition between
localization, favored by two-body repulsion, and delocalization,
favored by the hopping term leads to a quantum phase transition
between a `Mott insulator' phase and a `superfluid' phase in the
ground state as the relative importance of the two parameters $J$ and
$U$ of the model is varied~\cite{Fisher, Greiner}.

Specializing now to the case of a finite, one-dimensional potential
array consisting (for convenience) of an odd number of sites $M$ with
periodic boundary conditions (i.e., a ring-shaped one dimensional
array of sites), interesting symmetries become manifest by changing to
the representation which diagonalizes the hopping term of the
Hamiltonian (\ref{1}). This is achieved by a (discrete) Fourier
transform
\begin{equation}\label{2}
\Ahat_{q} \equiv \frac{1}{\sqrt{M}}\sum^{M}_{n = 1} e^{\frac{2\pi
    i}{M}nq}\ahat_n,
\end{equation}
where $q =-\frac{M-1}{2},\dots,0,\dots,\frac{M-1}{2}$. In
terms of the new `quasimomentum' bosonic operators $\Ahat_{q}$,
$\Ahat_{q}^\dagger$ the Hamiltonian reads
\begin{equation}\label{3}
\begin{array}{ccl}
H_{BH} &=&\displaystyle{ -2J \sum^{\frac{M-1}{2}}_{q =-\frac{M-1}{2}}
\cos(\frac{2\pi q}{M})\Ahat_q^{\dagger}\Ahat_q} \\ \\
&+& \displaystyle{\frac{U}{2M}\sum_{q_j}\delta_M(q_1 + q_2 - q_3 -
q_4)\Ahat^{\dagger}_{q_4}\Ahat^{\dagger}_{q_3}\Ahat_{q_2}\Ahat_{q_1}},
\end{array}
\end{equation}
where the $\delta_M(q)$ in the two-body term is the modular Kronecker delta,
equal to one if $q$ is an integer multiple of $M$ and zero
otherwise. It indicates `modular' conservation of total quasimomentum
($Q_T = \mod(\sum_qqn_q,\;M)$), i.e, with allowance for {\it Umklapp}
processes. This in fact reduces many-body Hamiltonian matrices to
block-diagonal form, each block being associated to a value of $Q_T$,
which assume the same values of $q$. In particular, traces of the
tight-binding 
assumption are manifest in this representation in the cosine law for
the single-particle energies and in the single common value $U/M$ for
the two-body matrix elements.

\subsection{Configuration space field theoretical model}

Limitations of the tight binding regime which are built into the
Bose-Hubbard Hamiltonian can be lifted without substantially
increasing eventual computational costs by first returning to a
configuration space representation of the Hamiltonian. We do this by
keeping the overall geometry consisting of a regular toroidal array
with mean radius $R$ and containing $M$ angular domains separated by
potential barriers, with a one-dimensional lattice constant given by
$l_c = 2\pi R/M$. Furthermore, we assume an effective one-dimensional
regime in which the transverse amplitude is independent of angle and
effectively frozen in its ground state. In this context the relevant
Hamiltonian can be written in terms of angle dependent bosonic field
operators $\psihat (\varphi)$, $\psihat^{\dagger}(\varphi)$
as~\cite{PethickBook, PitaevskiiBook} 
\begin{equation}
\begin{array}{cr}
H = &\displaystyle{\int d\varphi\;\psihat^{\dagger}(\varphi)
\Bigg[\left(-\frac{\hbar^2}{2mR^2}\frac{d^2}{d\varphi^2}+V_{latt}
(\varphi)\right)+}\\  
 &\displaystyle{+\frac{\Lambda}{2}\psihat^{\dagger}(\varphi)
\psihat(\varphi)\Bigg]\psihat(\varphi),}   
\end{array}
\label{Hnow}
\end{equation}
in which $m$ is the boson mass and $V_{latt}(\varphi)$ is the external
lattice potential, accounting both for tight transverse confinement
and for the angular periodicity. The effective one-dimensional
strength parameter $\Lambda$ is related to the strength of the usual
effective two-body contact interaction given in terms of the
$s$-wave scattering length $a$, $\lambda = 4\pi\hbar^2a/m$, as
\[
\Lambda=\lambda\int d^2r_\perp\;|w_0(\vec{r}_\perp)|^4.
\]
Here $w_0(\vec{r}_\perp)$ is the frozen transverse amplitude in the
array. This is a nodeless, normalized, confining wavefunction which
sets the scale for the transverse size of the toroidal trap, and thus
also the proportionality constant relating the effective strength
parameters $\lambda$ and $\Lambda$.

Before re-establishing contact with the Bose-Hubbard form~(\ref{3}) of
the Hamiltonian, we consider a further extension of the effective
one-dimensional Hamiltonian~(\ref{Hnow}) to include uniform rotation
of the lattice with angular velocity $\omega$ around the axis of the
toroidal structure. By transforming to the reference frame rotating
with the lattice potential, the required effective Hamiltonian becomes
\begin{equation}\label{4}
\begin{array}{ccc}
H_{\omega} &=& \displaystyle{\int d\varphi
\;\psihat^{\dagger}(\varphi)\Bigg[\frac{1}{2mR^2}\left(\frac{\hbar}{i}
\frac{d}{d\varphi}-m\omega R^2\right)^2 + }\\\\ &&\displaystyle{
+V_{\rm latt}(\varphi) +\frac{\Lambda}{2}\psihat^{\dagger}(\varphi)\psihat(
  \varphi)\Bigg]\psihat(\varphi).}
\end{array}
\end{equation}
This differs from Eq.~(\ref{Hnow}) just by the replacement
\begin{equation}
l_z\equiv\frac{\hbar}{i}\frac{d}{d\varphi}\longrightarrow
(l_z-m\omega R^2)\equiv\left(\frac{\hbar}{i}\frac{d}{d\varphi}-
m\omega R^2\right).
\label{3andahalf}
\end{equation}

One can now make contact with the Bose-Hubbard form~(\ref{3}) by
representing $H_\omega$ in the (truncated) single particle basis
consisting of the $M$ (first band) Bloch functions
$\{\phi_q^{(\omega)}(\varphi)\}$ which diagonalize its one-body
part. Note that these functions are labeled by the quasimomentum $q$
and depend on the cranking angular velocity $\omega$. This is done by
expressing $H_\omega$ in terms of the bosonic operators
\[
\begin{array}{cccc}
\Ahat_q^{\dagger(\omega)} &=&
\displaystyle{\int
  d\varphi\;\phi_q^{*(\omega)}(\varphi)\psihat^{\dagger}_q(\varphi)}
&\;\;\text{and}\\
\Ahat_q^{(\omega)} &=& \displaystyle{\int
d\varphi\;\phi_q^{(\omega)}(\varphi)\psihat_q(\varphi)},&  
\end{array}
\]
which leads to
\begin{equation}\label{5}
\begin{array}{lll}
H^{(\omega)} & = & \displaystyle{\sum^{M-1}_{q =
    0}e_q^{(\omega)}\Ahat_q^{(\omega)\dagger}\Ahat_q^{(\omega)}+} \\ 
 & &\displaystyle{ +\frac{\Lambda}{2}\sum_{q_i}{g}^{(\omega)}_{\{q_i\}}
  \delta_M(q_3 + q_4 - q_1 - q_2)\times} \\ 
& & \hspace{2cm}
\times\Ahat^{(\omega)\dagger}_{q_1}\Ahat^{(\omega)\dagger}_{q_2}
\Ahat^{(\omega)}_{q_3}\Ahat^{(\omega)}_{q_4}.  
\end{array}
\end{equation}
Here the $e_q^{(w)}$ are the (first band) single particle Bloch
eigenvalues and the objects ${g}^{(\omega)}_{\{q_i\}}\equiv
{g}^{(\omega)}_{q_1,q_2,q_3,q_4}$ are integrals over products of
four Bloch functions which, together with the strength parameter
$\Lambda$, constitute the required two-body matrix elements in the
adopted single-particle basis. The single particle energies are now
the Bloch eigenvalues which, as shown in continuation, have spacings
consistent with the cosine rule of~(\ref{3}) in the tight binding
regime; the two body matrix elements may and indeed do increasingly
show fluctuations as one leaves the extreme tight binding limit.

Finally, it is worth mentioning explicitly that truncation of the
single-particle basis to the first band Bloch states, together with
the modular conservation of total quasi-(angular)momentum in~(\ref{5})
implies that there are no `mean field two-body effects, i.e. all
two-body matrix elements involving just one-particle changes of state
vanish. As a consequence, reduced one-body density matrices of
non-degenerate stationary many-body states of~(\ref{5}) will be
diagonal in this base, total quasimomentum being then a good quantum
number.

\subsection{Numerical implementation}

\begin{figure}
    {\label{fig:cruz100}\includegraphics[width=3.4in]{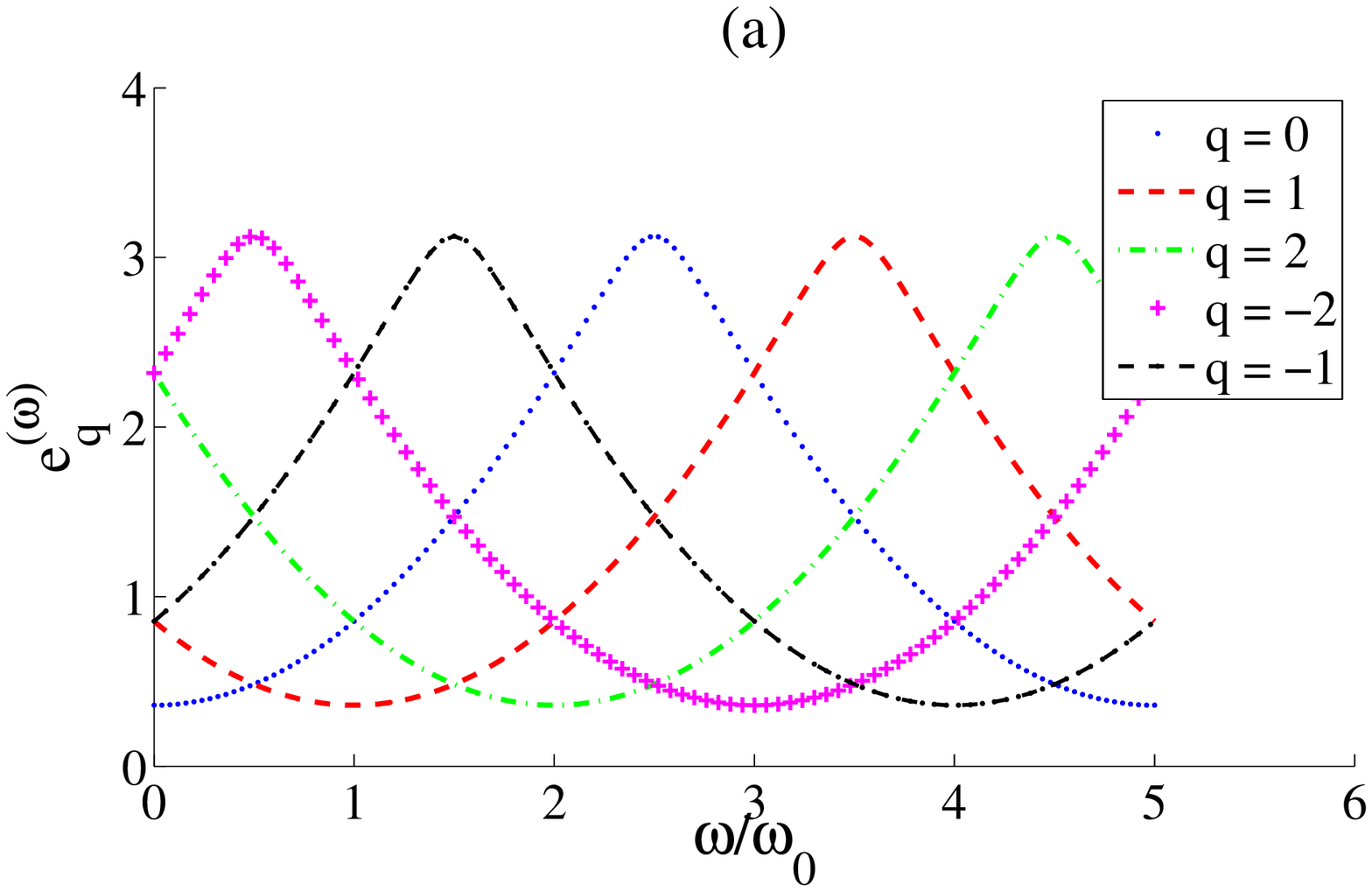}}
    {\label{fig:blahblah}\includegraphics[width=3.4in]{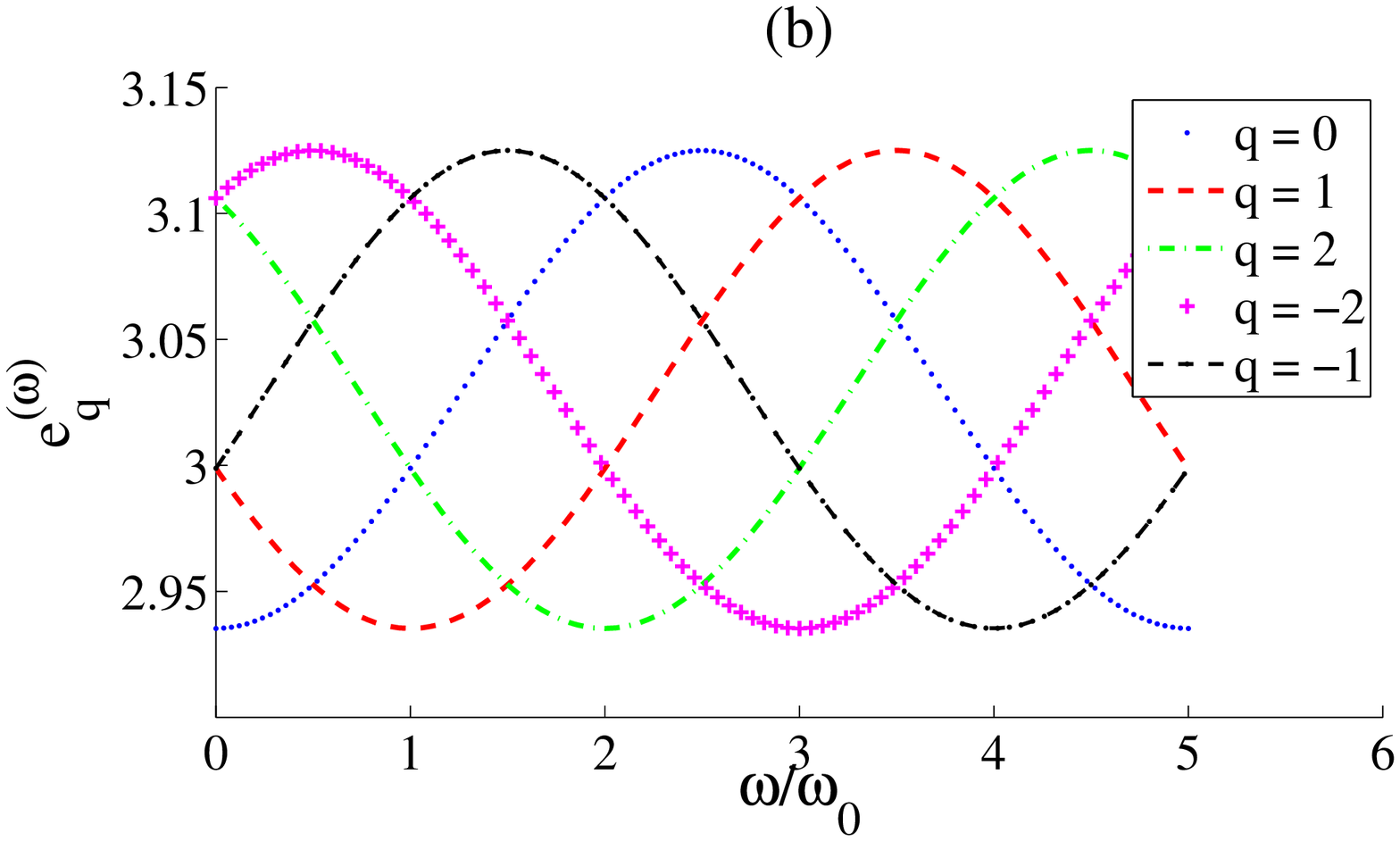}}
    {\includegraphics[width=3.4in]{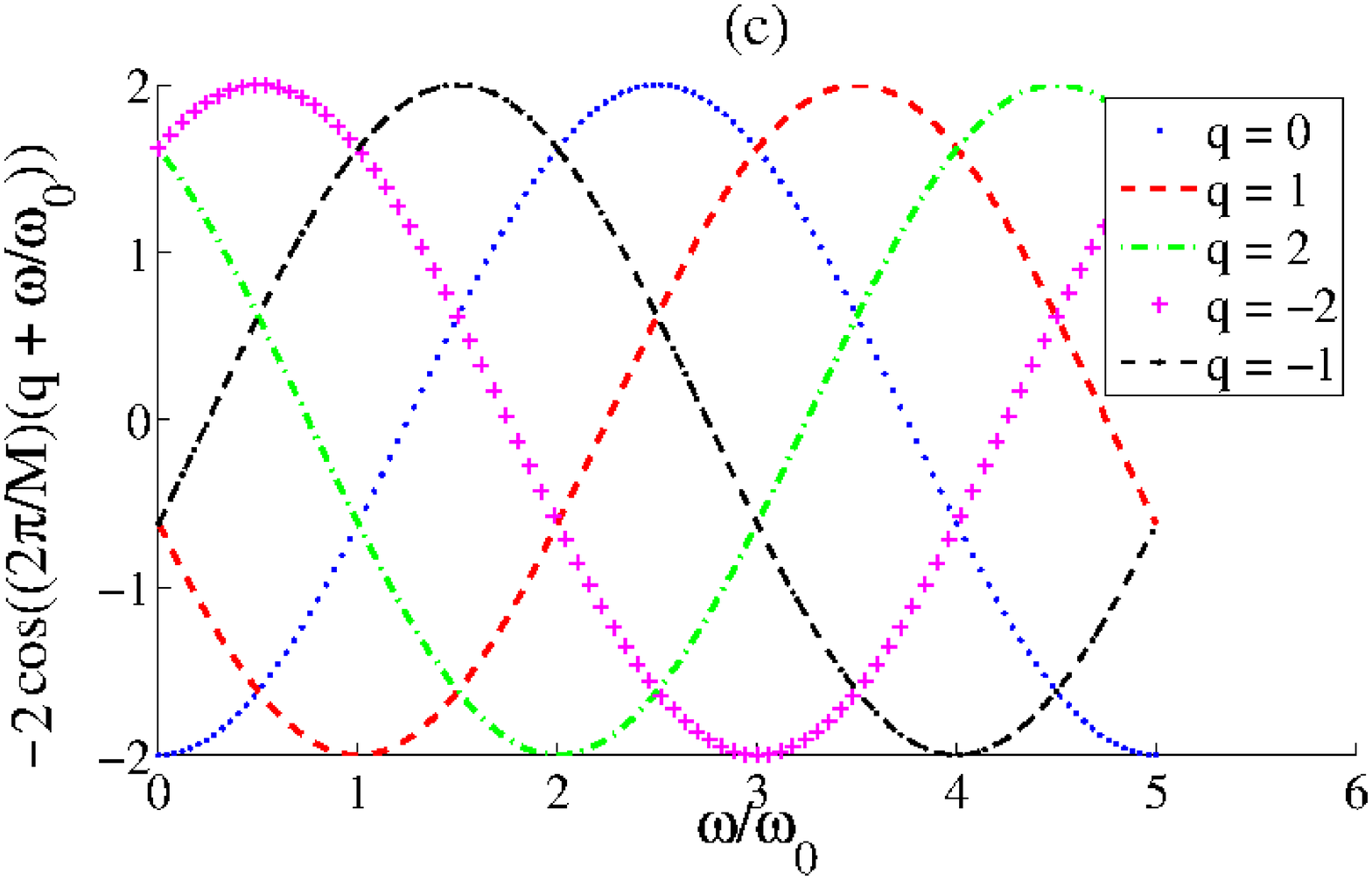}}
    \caption{(Color online) First band Bloch single-particle energies
      (in units of $\epsilon_0=\hbar^2/mR^2$), for $M = 5$, $\gamma =
      1\epsilon_0$ (a) and $100\epsilon_0$ (b), as functions of the
      cranking angular velocity $\omega$ (in units of $\omega_0$) (c):
      Bose-Hubbard single-particle energies per unit hopping parameter
      also as a function of $\omega$. Note the distortion relative to
      the cosine law (c) in case (a). One quarter of the width of the
      Bloch band may be used as an effective Bose-Hubbard hopping
      parameter $J$ for each value of $\gamma$, as discussed in the
      text. The values of $J$ for (a) and (b) are $0.691$ and $0.0474$
      respectively, also in units of $\epsilon_0$.}
\label{fig:bunching}
\end{figure}

\begin{figure}
\begin{center}
\includegraphics[scale=0.4]{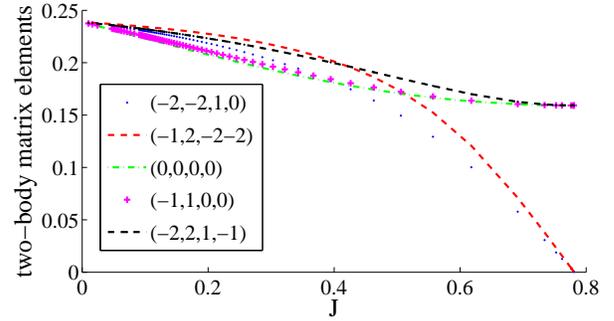}
\end{center}
\caption{(Color online) Typical behavior of the two-body matrix
  ele\-ments. The   notation $(q_1,q_2,q_3,q_4)$ denotes the integral
  $\int d\varphi 
  \phi^{*}_{q_1}(\varphi)\phi^{*}_{q_2}(\varphi)\phi_{q_3}(\varphi)\phi_{q_4} 
  (\varphi)$. Unlike strictly quasi-mo\-men\-tum conserving matrix 
  elements, {\it Umklapp} matrix elements vanish together with  
  barrier strength parameter at $J=25/32\epsilon_0=0.78125\epsilon_0$. } 
\label{fig:umklapp}
\end{figure}

In order to implement the many-body model~(\ref{5}) numerically we
first adopt a schematic realization of the lattice potential in terms
of the Kronig-Penney model with $\delta-$function barriers
\[
V_{\rm latt} \rightarrow \gamma \sum^{M-1}_{n = 0}\delta(\varphi -
\frac{2\pi}{M}(n + 1)),
\]
where $\gamma$ is the strength parameter for the $\delta-$function
barriers. For this choice, Bloch functions are given as analytic
expressions involving few numerical parameters which are easily
obtained, together with the Bloch eigenvalues, by solving transcendent
algebraic equations numerically.

The behavior of the Bloch eigenvalues as a function of the cranking
angular velocity $\omega$, given in units of $\omega_0 = \hbar/mR^2$
(energy being measured in units of $\epsilon_0\equiv\hbar^2/mR^2$), is
shown for two values of the barrier parameter $\gamma$ in
Fig.~\ref{fig:bunching} (a) and (b). Part (c) of the same figure shows
the 
$\omega$ dependence of the tight-binding regime one-body eigenvalues
as implemented in the Bose-Hubbard model, given by the function
$-2\cos(\frac{2\pi}{M}(q-\omega))$ (cf. Eq.~(\ref{3})). While this
cosine function is clearly capable of reproducing the relative
spacings of the Bloch single-particle energies very accurately for
sufficiently large values of $\gamma$, deviations from it are clearly
seen in the case of the lower value of $\gamma$. We use single
particle Bloch eigenvalues $e_q^{(w)}$ obtained numerically in all
calculations, while taking advantage of the analytical expression
valid in the tight binding regime to {\it define} a parametrization of
the barrier strength in terms of an effective hopping coefficient $J$
as
\begin{equation}
J\equiv\frac{\epsilon_{\gamma}^>-\epsilon_{\gamma}^<}{4}.
\label{Jdef}
\end{equation}
Here $\epsilon_{\gamma}^>$ and $\epsilon_{\gamma}^<$ are respectively
the upper and the lower bounds for the first band cranked Bloch
eigenvalues $\{\epsilon_q^{(\omega)}\}$ for a given value of the
barrier strength parameter $\gamma$. For simplicity we refer to the
numerator of Eq.~(\ref{Jdef}) as the `energy width of the first
band'. This parametrization coincides with the Bose-Hubbard definition
of the hopping parameter in Eqs.~(\ref{1}) and~(\ref{3}) in the
tight-binding domain. Note also that the dependence of the Bloch
energies with the cranking angular velocity leads to single-particle
energy level crossings at integer and and half-integer values of
$\omega/\omega_0$. Since the quasimomentum is a good quantum number
for the individual Bloch states, these level crossings will have a
decisive role in determining the total quasimomentum of cranked
many-body ground state, as discussed in the following section.

The four wavefunction integrals
${g}^{(\omega)}_{q_1,q_2,q_3,q_4}$ are easily evaluated in terms
of the Bloch functions. Some sample results are shown in
Fig.~\ref{fig:umklapp}, for $M=5$ and $\omega=0$, as functions of the
effective hopping parameter $J$ just defined. Salient features here
are the complete bunching on a single value ${g}_0$ in the
tight-binding limit $J\rightarrow 0$ (${g}_0\sim 0.239$ in the
present case), and the strong quenching of {\it Umklapp} matrix
elements in the opposite limit of very large hopping, as compared to
angular momentum conserving matrix elements. Note that the
definition~(\ref{Jdef}) of the effective hopping parameter leads to 
$J/\epsilon_0\rightarrow 25/32=0.78125$ as the barrier strength
parameter $\gamma\rightarrow 0$, for the present case with $M=5$. In
order to facilitate comparison with results obtained using the
Bose-Hubbard Hamiltonian parameters as in~(\ref{1}) and~(\ref{3}), we
again use the tight-binding limit to {\it define} an effective
two-body parameter $U$ to replace the two-body constant $\Lambda$ as
\begin{equation}
U\equiv{g}_0M\Lambda
\label{Udef}
\end{equation}
in terms of which the two-body part of Eq.~(\ref{5}) coincides in the
tight-binding limit to that of the Bose-Hubbard model, Eq.~(\ref{3}).

\section{The ground-state phase diagram}\label{sec:phasediagram}

\begin{figure}[htp]
\includegraphics[scale=0.34]{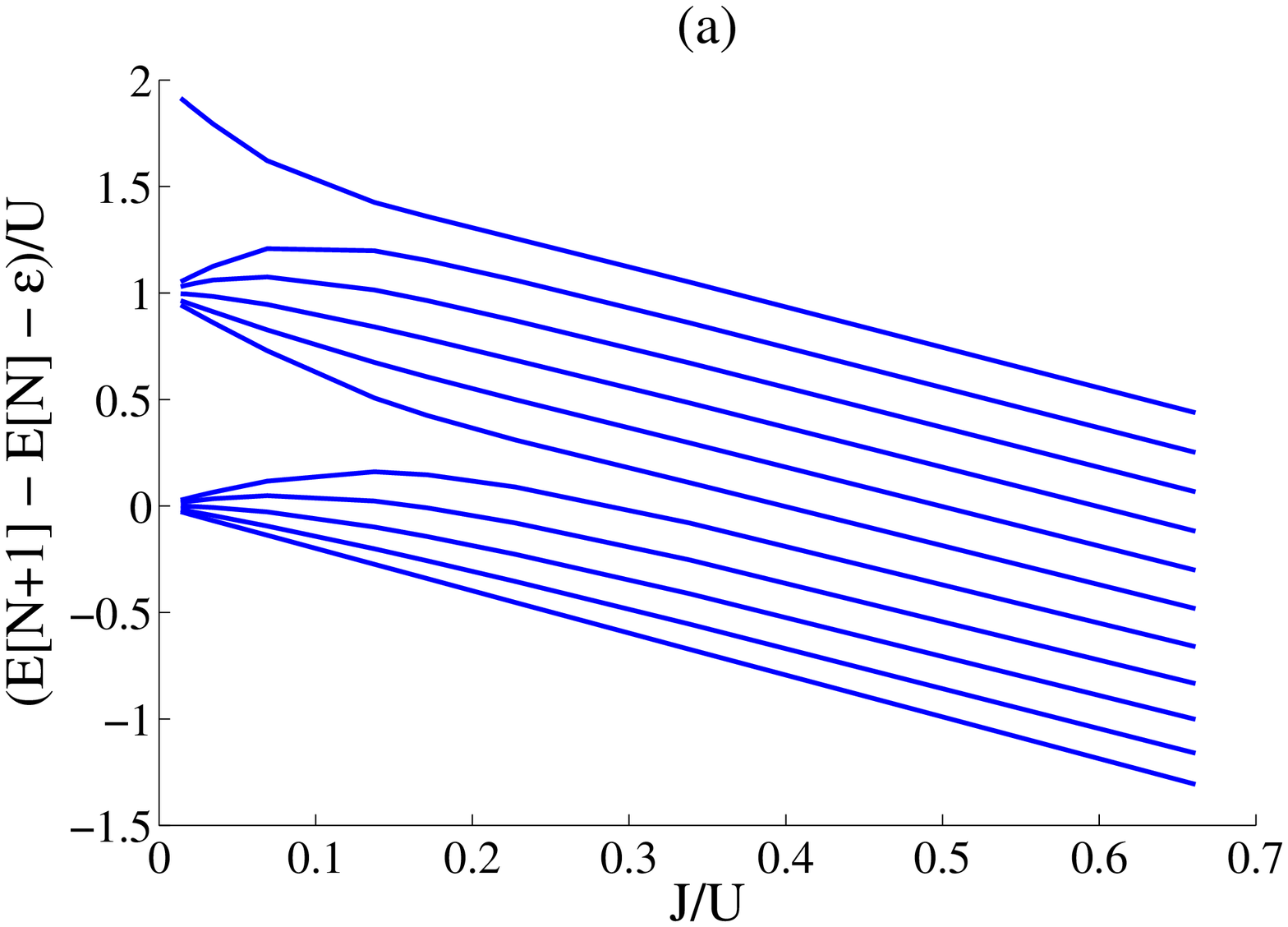}
\includegraphics[scale=0.34]{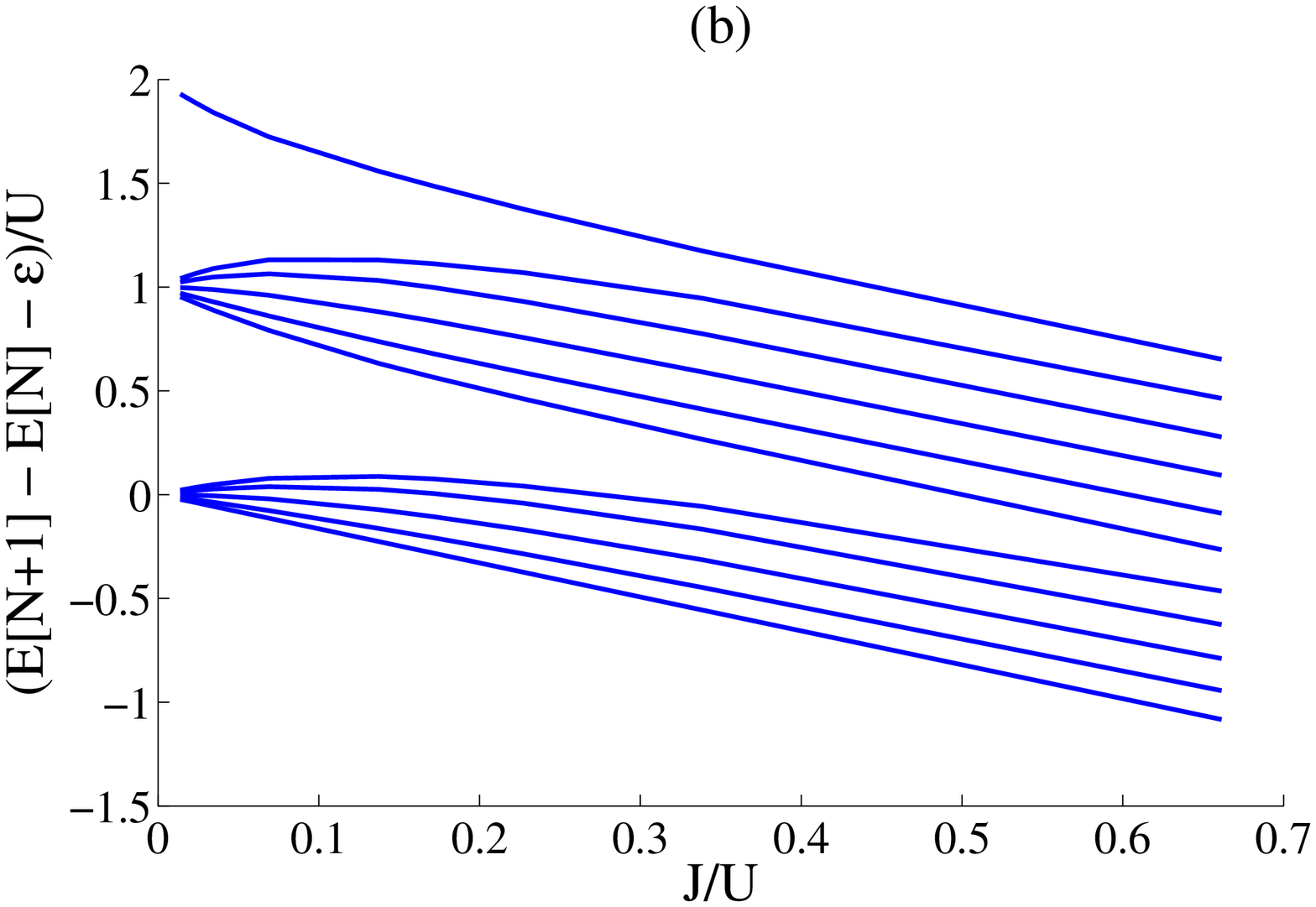}
\caption{Scaled particle addition energies as functions of $J/U$ for
  $\omega = 0$ and $\omega = 0.48\omega_0$ using (the) single-particle
  energies, and with the effective hopping parameter defined in the
  text. Here $\Lambda = 0.06\epsilon_0$, $M = 5$ and $N$ varies from 0
  to 11 (from the first to the last line going upwards). In (a) the
  closure of the lowest Mott lobe is seen to happen in the
  neighborhood of $J/U \sim 0.35$, while in (b) it is affected by the
  angular velocity and happens for higher values, $J/U > 0.7$.}
\label{fig:mottlobes}
\end{figure}

The different dynamical regimes prevailing in different parameter
domains~\cite{Fisher} of the Bose-Hubbard Hamiltonian~(\ref{3}) have
been studied extensively using diverse approximation schemes and/or
computational techniques.  
The focus of these studies is the
thermodynamic limit, in which the number of sites and the number of
bosons go to infinity at fixed finite mean occupation per site. The
one-dimensional case has been treated rather recently by K\"{u}hner, White
and Monien~\cite{KWM}, including, in particular, the `ground state phase
diagram'~\cite{Fisher} in which different phase domains are identified
in a $\mu/U\times J/U$ diagram, $\mu$ being the chemical potential.

Our purpose in this section is to indicate how one can obtain
phase-diagram information by using the results of exact many-body
diagonalization of the Hamiltonian~(\ref{5}), albeit with feasibly
small number of sites and of bosons. Under these circumstances there
will of course be no case for taking thermodynamic limits and
determining precise phase boundaries, but as will be shown explicitly
there are clear `precursor features' in the small number solutions
which clearly identify, surprisingly even quantitatively, several
thermodynamic limit properties. 

Since in this context we deal always with systems having a fixed,
given number of particles, a replacement must be devised for the
chemical potential $\mu$. We thus replace $\mu$ by the addition energy
$\Delta_N$, $N$ being the number of bosons, defined as $\Delta_N=
E_0(N+1)-E_0(N)-\epsilon$, where $E_0(N)$ denotes the ground
state eigenvalue for $N$ bosons in the chosen number of sites, and
$\epsilon$ is the average of the single particle (first-band
Bloch) energies. Note that $\epsilon$ vanishes for the cosine
law of the Bose-Hubbard Hamiltonian~(\ref{3}). With this replacement,
the axes of the graph corresponding to the ground state phase diagram
become
\begin{equation}
\frac{\mu}{U}\rightarrow\frac{\Delta_N}{U}=\frac{E_0(N+1)-E_0(N)-
\epsilon}{U},\hspace{.5cm}\frac{J}{U}
\label{Ngspd}
\end{equation}
where we use the quantities $U$ and $J$ as defined in
Eqs.~(\ref{Udef}) and~(\ref{Jdef}) in order to characterize the
dynamical parameters of the many-body Hamiltonian~(\ref{5}).

Results obtained for $M=5$ sites and $N=0$ to $11$ bosons are shown in
Fig.~\ref{fig:mottlobes} (a), for the usual situation of vanishing
cranking angular velocity, $\omega=0$, the strength of the two-body
effective interaction having been fixed at $\Lambda=0.06
\epsilon_0$. This value is realistic in the sense that it
corresponds to the mass and scattering length of $^{87}$Rb for $R\sim
10\;\mu$m with a transverse confinement scale of the order of
$1\;\mu$m. The value of the barrier strength parameter $\gamma$ has
been varied to cover the desired range of values of $J/U$. Actually
the whole parameter domain covered by this graph falls within the
tight-binding domain in which the results obtained using the
Hamiltonian~(\ref{5}) differ very little from what one obtains using
the Bose-Hubbard Hamiltonian~(\ref{3})
itself~\cite{PizaDiagrams}. Each curve in this graph shows the
dependence on $J/U$ of the ($U$-scaled) addition energy $\Delta_N$ for
one of the values of $N$. As seen, the bunching of these curves at
integer values of the $U$-scaled addition energies in the limit
$J/U\rightarrow 0$ gives room for the Mott Insulator lobes. The shape
and range of these lobes reproduces even quantitatively the results
obtained towards the thermodynamic limit, cf. Ref.~\cite{KWM},
including in particular the `reentrant behavior'~\cite{PPC} (see also
Ref.~\cite{Giamarchi}) characteristic of the one-dimensional
Bose-Hubbard model. Moreover, one can show perturbatively, in the
large $J/U$ limit past the insulator lobes, that the spacing between
consecutive $\Delta_N/U$ curves approaches the value $1/M$, $M$ being
the number of sites~\cite{PizaDiagrams}, suggesting therefore a
definite $N$-cleavage as one moves toward the thermodynamic limit.

Effects of cranking are illustrated in Fig.~\ref{fig:mottlobes} (b),
which differs from (a) in that here $\omega=0.48\omega_0$, i.e. just
shortly before of the first single-particle level crossing (see
Fig.~\ref{fig:bunching}). The broadening and lengthening of the Mott
insulator lobes which is visible in this case is governed essentially
by the Bloch energies modified by rotation. This effect can be
understood in terms of the effective reduction of the kinetic energy
(cf. Eq.~(\ref{3andahalf})) which quenches hopping thus favoring the
insulating phase.
 
\section{Properties related to superfluidity}\label{sec:superfluidity}

The dependence on $\omega$ of the energy (in the rotating system) of
the many-body ground state of the cranked Hamiltonian~(\ref{3}), for
small values of the angular velocity, allows for the determination of
an inertial parameter ${\cal{I}}$ through the relation
\[
E^{(\omega)}(N)=E^{(0)}(N)+\frac{\cal{I}}{2}\omega^2+
{\cal{O}}(\omega^4),\hspace{.5cm}\omega\rightarrow 0
\]
which may be related to the `superfluid fraction' $f_s$,
defined within the phenomenological framework of the two-fluid model
of superfluid behavior, as (cf. Ref.~\cite{LiebBook})
\begin{equation}
f_s=\frac{\cal{I}}{\cal{I}}_{\rm Rig}=\frac{\cal{I}}{NmR^2},
\label{fs_I}
\end{equation}
where ${\cal{I}}_{\rm Rig}=NmR^2$ is the rigid moment of inertia of
the system. This bears the understanding that, in stationary state,
as viewed from the rotating frame, the inertia corresponds to that
part of the fluid which remains stationary in the lab frame (the
`superfluid'), while the `normal' component is carried around with the
externally imposed angular velocity.

\subsection{Current for inertial parameter} 

The inertial parameter ${\cal{I}}$ can be expressed as
\begin{equation}
{\cal{I}}=\left.2\frac{dE^{(\omega)}(N)}{d\omega^2}\right|_{\omega
\rightarrow 0}=\left.\frac{1}{\omega}\langle\Phi^{(\omega)}|
\frac{dH_\omega}{d\omega}|\Phi^{(\omega)}\rangle\right|_{\omega
\rightarrow 0}
\label{FHI}
\end{equation}
where $|\Phi^{(\omega)}\rangle$ is the state vector for the $N$-body
ground state at cranking angular velocity $\omega$ and where the well
known Feynman-Hellmann relation has been used. This last expression
involves the expectation value of the derivative of $H_\omega$,
Eq.~(\ref{4}), with respect to the parameter $\omega$. This derivative
is in fact a one-body (albeit nonlocal) operator:
\begin{equation}
\frac{dH_\omega}{d\omega}=-\int d\varphi\,\psihat^\dagger(\varphi)
\left(\frac{\hbar}{i}\frac{d}{d\varphi}-mR^2\omega\right)
\psihat(\varphi).
\label{dHdomega}
\end{equation}

All the information relevant for one-body observables, contained in
the correlated many-body ground state $|\Phi_0^{(\omega)}\rangle$ of
the Hamiltonian~(\ref{5}), is carried by its associated one body
reduced density, defined as
$\rho^{(1)}(\varphi,\varphi')\equiv\langle\Phi_0^{(\omega)}|
\psihat^\dagger(\varphi')\psihat(\varphi)|\Phi_0^{(\omega)}\rangle$.
We may thus proceed to express the inertial parameter ${\cal{I}}$, as
given by Eq.~(\ref{FHI}), in terms of this object. For this purpose, it
is convenient to use its spectral decomposition, which involves the
solutions of the eigenvalue problem
\[
\int d\varphi^{'}\rho^{(1)}(\varphi, \varphi^{'})\chi_\nu(\varphi^{'}) =
n_\nu\chi_\nu(\varphi). 
\]
where the single-particle eigenfunctions $\chi_\nu(\varphi)$ are the
so called natural orbitals, the associated eigenvalues $ n_\nu$ being
the corresponding occupation numbers~\cite{Yukalov}. In terms of
these ingredients the one body reduced density can be expressed as
\begin{equation}
\rho^{(1)}(\varphi, \varphi^{'}) =
\sum_\nu n_\nu\chi_\nu(\varphi)\chi_\nu^* (\varphi^{'}).
\label{specd}
\end{equation}
Using~(\ref{dHdomega}) and~(\ref{specd}) in Eq.~(\ref{FHI}) one
finds that the inertial parameter $\cal{I}$ can be written in the form
\begin{equation}
{\cal{I}}=\left.\frac{mR^2}{\omega}\int d\varphi\,j_\omega
(\varphi)\right|_{\omega\rightarrow 0}
\label{Ifromj}
\end{equation}
where the current $j_\omega(\varphi)$ is given by
\begin{equation}
j_\omega(\varphi)\equiv-\sum_{\nu}n_{\nu}\left[\frac{\hbar}{mR^2}{\rm Im}
\left(\chi_{\nu}^*(\varphi)\frac{d\chi_{\nu}}{d\varphi}\right)-
\omega|\chi_{\nu} (\varphi)|^2\right].
\label{scurr}
\end{equation}
To obtain this expression, matrix elements involving a 
$\varphi$-derivative have been evaluated with the prescription
\[
\int d\varphi\,\psihat^\dagger(\varphi)\frac{d}{d\varphi}
\psihat(\varphi)=\int d\varphi\int d\varphi'\,\psihat^\dagger
(\varphi)\delta'(\varphi-\varphi')\psihat(\varphi')
\]
where $\delta'(\varphi)$ is the first derivative of the Dirac delta
function with respect to the argument.

The current $j_\omega(\varphi)$ is therefore, in general, an incoherent
sum of currents associated to each natural orbital, full coherence
resulting only in the limiting case of full occupancy of a single
natural orbital. Moreover, it is easy to see that the current
$j_\omega(\varphi)$, together with the (diagonal part of the) one-body
reduced density, obeys a conservation law which, for the stationary
states of the cranked Hamiltonian~(\ref{4}), in fact makes it {\it
  independent of $\varphi$} (i.e., solenoidal) as a result of the time
independence of the associated one-body reduced density.

In order to see this, we start from the expression of the many-body
Hamiltonian in terms of the field operators, Eq.~(\ref{5}), and use
the stationary character of $|\Psi^{(\omega)}\rangle$
\begin{eqnarray*}
H_\omega\vert\Psi^{(\omega)}(t)\rangle&=& i\hbar\frac{d}{dt}\vert
\Psi^{(\omega)}(t)\rangle,\\
\vert\Psi(t)^{(\omega)}\rangle&=&
\displaystyle{e^{-\frac{i}{\hbar}E_0^{(\omega)}t}
\vert\Phi^{(\omega)}\rangle} 
\end{eqnarray*}
to write the vanishing time derivative of the reduced one-body density
as
\begin{equation}
\begin{array}{l}
\displaystyle{\frac{d}{dt}\rho(\varphi,\varphi^{'})
\mid_{\varphi = \varphi^{'}} =\frac{d}{dt}\langle\Psi^{(\omega)}\vert
\psihat^{\dagger}(\varphi)  \psihat(\varphi^{'})  
\vert\Psi^{(\omega)}\rangle\mid_{\varphi = \varphi^{'}} =} \\\\
\displaystyle{\hspace{.5cm}=\frac{1}{i\hbar}\langle\Phi^{(\omega)}
\vert\left[\psihat^{\dagger}(\varphi)\psihat(\varphi^{'}),H_\omega
\right]\vert\Phi^{(\omega)}\rangle\mid_{\varphi = \varphi^{'}} = 0.} 
\end{array}\nonumber
\end{equation}
By computing the $\varphi$-derivatives with the prescription 
\begin{eqnarray*}
&&\int d\varphi\,\psihat^{\dagger}(\varphi)\frac{d^2}{d\varphi^2}
\psihat(\varphi) =\\
&=&\int d\varphi\int d\varphi^{'}\psihat^{\dagger}(\varphi)
\delta^{''}(\varphi - \varphi^{'})\psihat(\varphi^{'}), 
\end{eqnarray*}
where $\delta''(\varphi)$ stands for the second derivative of the
Dirac delta function with respect to the argument, the commutator can
again be expressed in terms of the reduced one-body density expressed
in terms of its natural orbitals with the result
\begin{equation}
\frac{d}{d\varphi}\sum_{\nu}n_{\nu}\left[\frac{\hbar}{mR^2}
{\rm Im}\left(\chi^{*}_{\nu}(\varphi)\frac{d\chi_{\nu}(\varphi)}{d\varphi}
\right)-\omega\vert\chi_{\nu}(\varphi)\vert^2\right] = 0,
\label{sol}
\end{equation}
which states the $\varphi$ independence of the current $j_\omega$.

Using~(\ref{sol}), the expression for the inertial parameter
${\cal{I}}$, Eq.~(\ref{Ifromj}), reduces to
\[
{\cal{I}}=\left.mR^2\frac{2\pi}{\omega}j_{\omega}\right|_{\omega\rightarrow 0}.
\]
Correspondingly, the `superfluid fraction' $f_s$ is expressed in terms
of this current as
\begin{equation}
f_s=\left.\frac{1}{N}\frac{2\pi}{\omega}j_\omega\right|_{\omega\rightarrow 0}.
\label{fs_j}
\end{equation}
In fact, the last two factors clearly amount to the time integral of
the solenoidal current $j_\omega$ for the duration of one period of
the cranking.

A feature of the evaluation of superfluid fractions by means of
Eqs.~(\ref{fs_I}) and~(\ref{fs_j}) which is worth stressing is that
the definition of the current $j_\omega$ stems directly from the
inertial parameter ${\cal{I}}$, being identified as the expectation
value of a momentum dependent one-body operator, and therefore apt to
be expressed in terms of the one-body reduced density, independently
of any assumptions concerning the relevance of a condensate
wavefunction~\cite{Keith2, Keith1}. This possibly manifests itself
{\it a posteriori}, through the coherence properties of the current.

It is worth noting that the formal
results~(\ref{Ifromj}),~(\ref{scurr}) and~(\ref{sol}) can be extended
in a straightforward way 
to a more general context, involving stationary many-boson states in
an arbitrary three-dimensional external trap, cranked around a fixed
axis specified by the unit vector $\vec{u}$. The effective
Hamiltonian~(\ref{4}) is in this case replaced by 
\begin{equation}\label{h3d}
\begin{array}{ccc}
H_{\omega} &\rightarrow& \displaystyle{\int d^3r
\;\psihat^{\dagger}(\vec{r})\Bigg[\frac{1}{2m}\left(\frac{\hbar}{i}
\vec{\nabla}-m\omega (\vec{u}\times\vec{r})\right)^2 + }\\\\ &&
\displaystyle{+V_{\rm trap}(\vec{r}) +\frac{\lambda}{2}\psihat^{\dagger}
(\vec{r})\psihat(\vec{r})\Bigg]\psihat(\vec{r}).}
\end{array}
\end{equation}
Following the procedure just described for the one-dimensional case,
one obtains from~(\ref{h3d}) an expression for the inertial parameter
that is given in terms of the volume integral of the appropriately weighted
tangential component of a solenoidal current density. This
current is again written as an incoherent sum of currents associated with
the natural orbitals, and that are weighted by the corresponding
occupation numbers.


\subsection{Numerical results}

For the case of the calculations reported here, the fact that the
many-body states $\vert\Phi^{(\omega)}_0\rangle$ have good total
quasi-momentum, together with the adopted truncation to single boson
states of the first band only, ensures that the associated reduced
one-body densities are diagonal in the quasi-momentum representation,
i.e.
\begin{equation}\label{b1}
\langle \Psi_0^{(\omega)}\vert \Ahat_q^{(\omega)\dagger}
\Ahat_{q'}^{(\omega)}\vert\Psi_0^{(\omega)}\rangle =n_q\delta_{qq'}. 
\end{equation}
The natural orbitals are thus just the Bloch functions
$\phi_q^{(\omega)}$ themselves, so that the corresponding occupation
numbers $n_q$ may accordingly be labelled by the associated
quasimomentum $q$. As a numerical check, we evaluate the superfluid
fraction $f_s$ both by taking numerical derivatives of the ground
state eigenvalue $E^{(\omega)}(N)$ with respect to $\omega^2$, as
written in Eq.~(\ref{FHI}), and in terms of the current $j_\omega$, as
in Eq.~(\ref{fs_j}), the current being evaluated as in
Eq.~(\ref{scurr}), in terms of the Bloch functions and of the
eigenvalues $n_q$ of the reduced one-body density, Eq.~(\ref{b1}). In
order to deal with the implied limit $\omega\rightarrow 0$ in the
numerical evaluation of superfluid fraction values from either
Eqs.~(\ref{fs_I}) and~(\ref{FHI}) or from Eq.~(\ref{fs_j}), we use the
fact that the value obtained is very stable against variation of the
value of $\omega$. This remains true even by orders of magnitude down
from $\omega\sim 0.1\omega_0$, as long as the differences involved
in obtaining either ${\cal{I}}$ or $j_\omega$ are sufficiently above
the limitations set by machine precision. In view of
Eq.~(\ref{Ifromj}), this amounts to a numerical verification that the
$\omega$-dependence of $E^{(\omega)}(N)-E^{(0)}(N)$ is quadratic,
while that of $j_\omega$ is linear within such a range.

\begin{figure}
    \label{fig:g4}\includegraphics[width=3.4in]{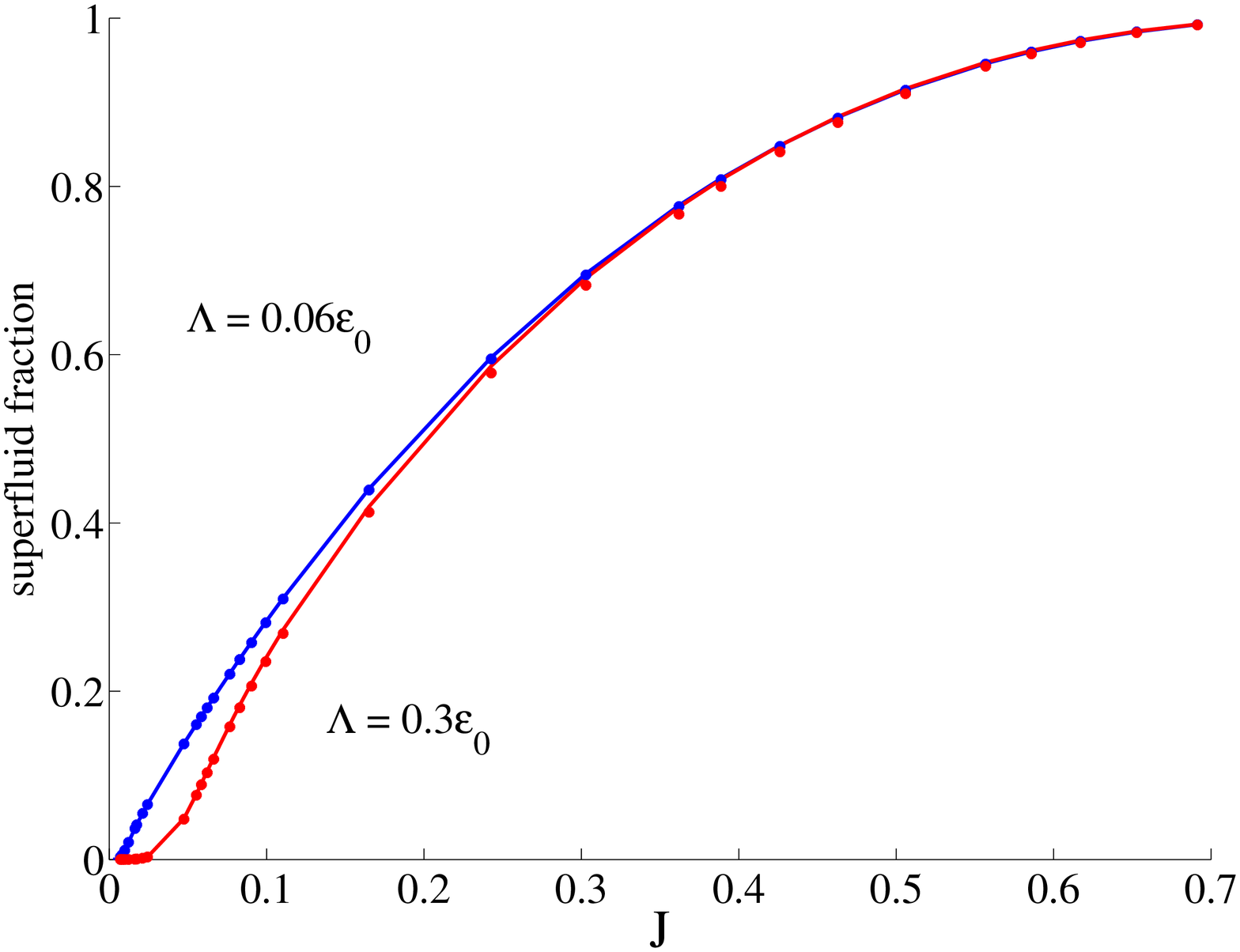}
    \caption{(Color online) Superfluid fraction for the system with $N
      = M = 5$ and 
      $\Lambda = 0.06\epsilon_0\;\,{\rm and}\;\,\Lambda = 0.3
      \epsilon_0$. The curves 
      correspond to Eqs.~(\ref{fs_I}) and (\ref{FHI}) while the points
      are obtained using the solenoidal current as in
      Eq.~(\ref{fs_j}). See text for details. }
\label{fig:current}
\end{figure}

Results for the commensurate case $N=M=5$ are shown in
Fig.~\ref{fig:current}., where the values obtained for the superfluid
fraction $f_s$ are plotted as a function of the hopping parameter $J$
for two values of the two-body strength parameter $\Lambda$ (or $U$,
see eq. (\ref{Udef})).  The most striking result of the numerical
evaluation of the superfluid fraction for the model system on hand is
the `prima facie' absence of any dramatic features in the parameter
domain ($J\sim 0.025\epsilon_0$ for $\Lambda=0.06\epsilon_o$ and
$J\sim 0.125\epsilon_0$ for $\Lambda=0.3\epsilon_0$) which supports
the transition associated with the closing of the Mott insulator lobes
(cf. Figs.~\ref{fig:mottlobes} (a) and~\ref{fig:current}). In fact,
while being affected by the increase of the two-body interaction
parameter, the superfluid fraction retains there a smooth
monotonically increasing behavior saturating at the full scale of the
hopping parameter $J$.

\begin{figure}
    \includegraphics[width=3.4in]{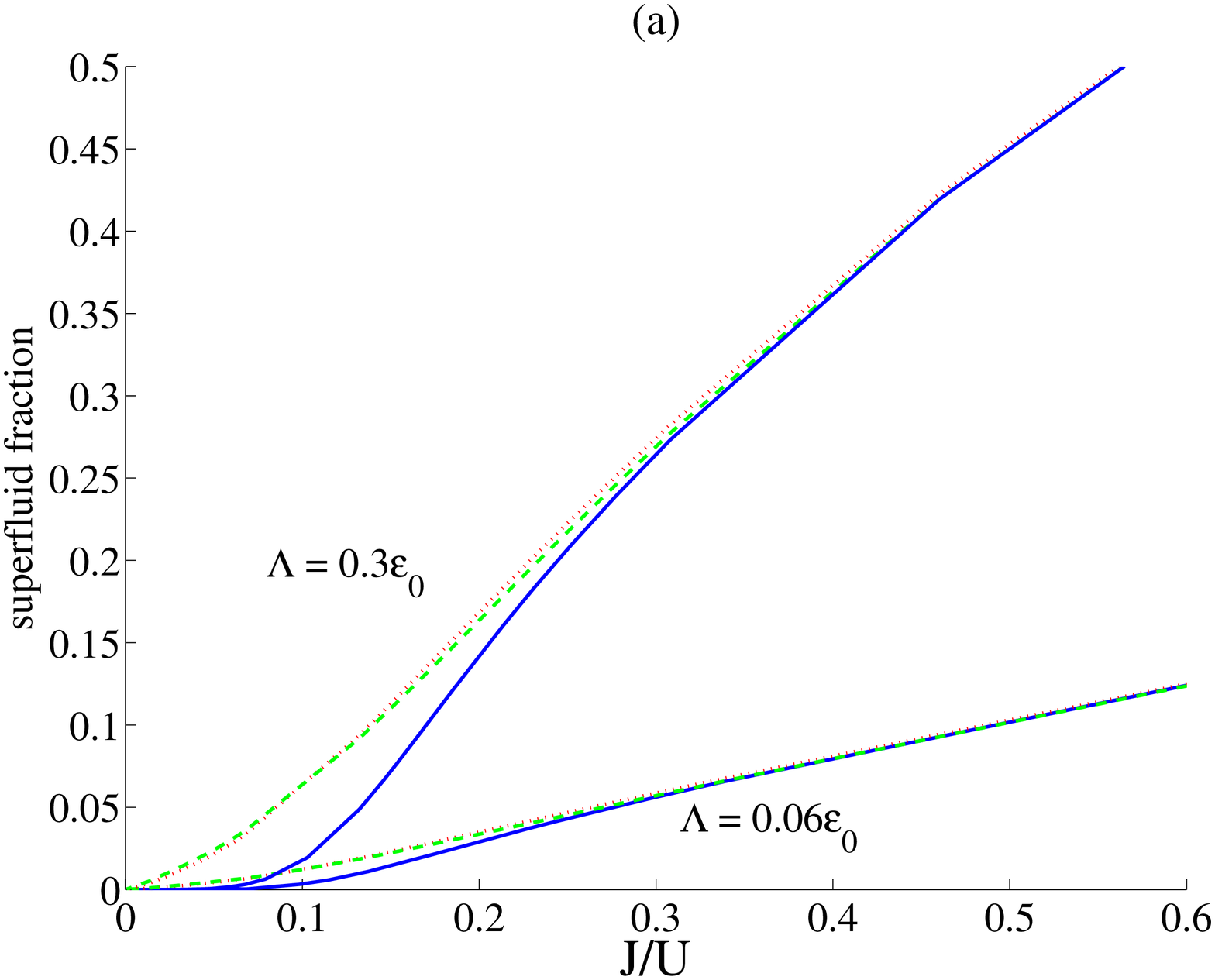}
    \includegraphics[width=3.4in]{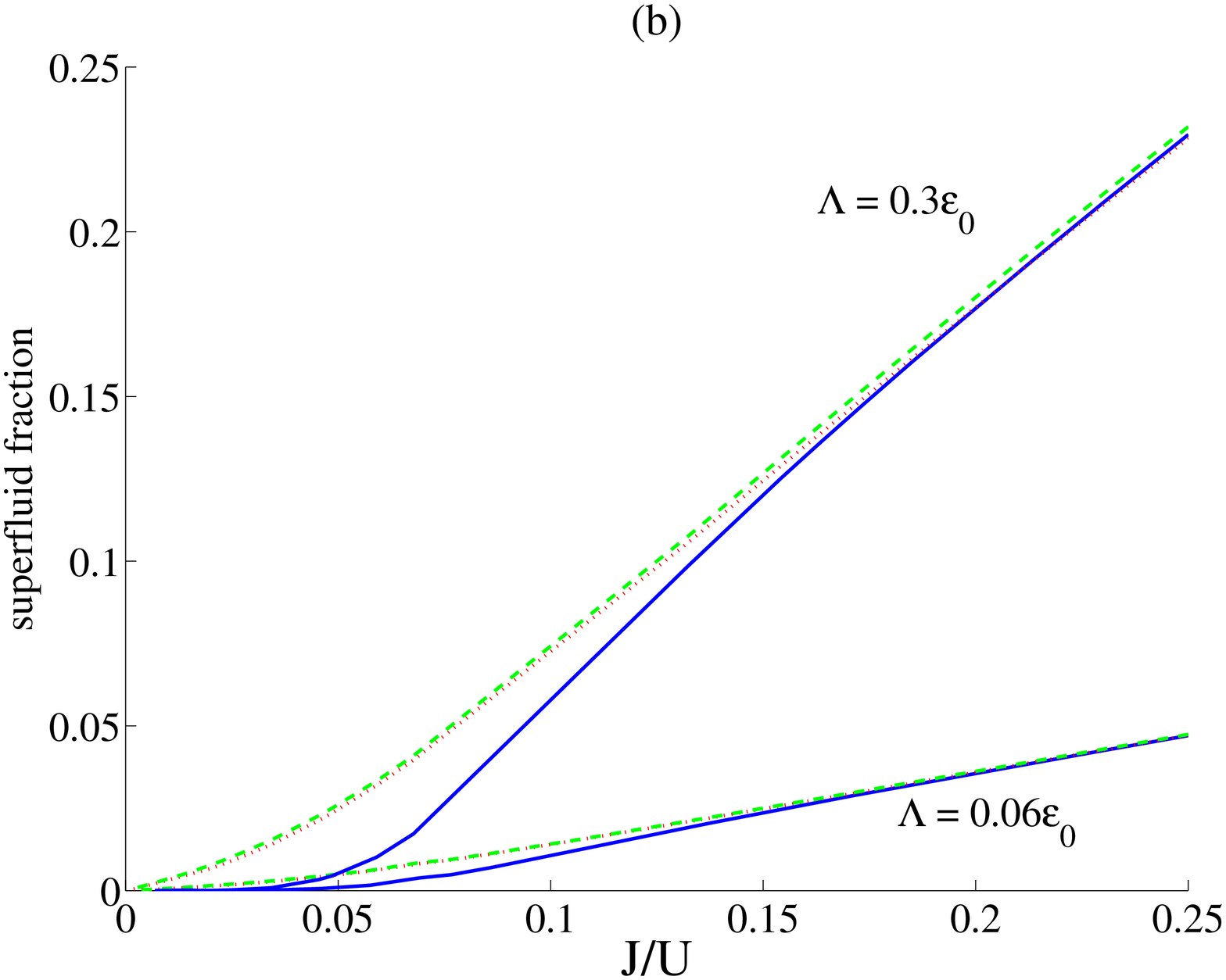}
    \caption{(Color online) Quenching of the superfluid fraction at
      commensurate fillings, $N_1=M$ in (a), $N_2=2M$ in (b) In both
      cases results are given for the two values
      $\Lambda=0.06\epsilon_0$ and  $\Lambda=0.3\epsilon_0$ of the two
      body strength parameter.  Curves for 
      $N_i-1$, $N_i$ and $N_i+1$ are shown, the lowest corresponding
      to commensurate filling. Compare the extension of the quenching
      interval of the variable $J/U$ to the extension of the
      corresponding Mott insulator lobes in Fig.~\ref{fig:mottlobes}
      (a).}
\label{fig:suplobes}
\end{figure}

Effects related to the presence of the Mott insulator lobes in the
ground-state phase diagram of Fig.~\ref{fig:mottlobes} (a) do appear,
however, in the corresponding intervals of $J/U$ and consist of a
quenching of the calculated superfluid fraction in such intervals at
the corresponding commensurate filling, relative to the values
obtained for incommensurate filling. This is shown in
Figs.~\ref{fig:suplobes} (a) and (b) for the cases $N/M=1$ and $2$
respectively. As discussed in Section~\ref{sec:condensation} below,
fragmentation of the total boson number $N$ over occupations $n_q$ of
different natural orbitals occurs in the $J/U$ domains spanned by the
Mott insulator lobes and is stronger in the case of commensurate
occupation (see Fig.~\ref{fig:occ} (a) and (d)). The quenching effect
may thus be associated with stronger loss of coherence of the current
$j_\omega$ (see Eq.~(\ref{Ifromj})) for commensurate filling. Note
that, for the `realistic' value $\Lambda=0.06\epsilon_0$ of the two body
strength parameter, the calculated values of the superfluid fraction
are small in the whole $J/U$ domains associated with the Mott
insulator lobes, so that these effects are not conspicuous on the
scale used in Fig.~\ref{fig:current}.

A comment relating the results obtained for the superfluid fraction as
understood in Ref.~\cite{Keith1} is also in order. A connection to the 
present evaluation of superfluid fractions can in fact be established
analytically in a rather straightforward way by noting that
Eq.~(\ref{FHI}) can in particular be applied to an ideal gas ($U=0$),
in which case the ground state for small $\omega$ consists of the $N$
bosons occupying the lowest ($q=0$) quasimomentum Bloch orbital. In
the tight binding approximation the ground state energy is thus given
by 

\[
E^{(\omega)}_{\rm ideal}(N)=-2JN\cos\left(\frac{2\pi\omega}{M\omega_0}
\right),\hspace{1cm}\omega_0=\frac{\hbar}{mR^2},
\]

\noindent from which one easily calculates the corresponding
superfluid fraction $f_s^{\rm (ideal)}$ in terms of
$dE^{(\omega)}_{\rm ideal}(N)/d\omega^2$. One can then set up a
quenching factor $Q$ defined as the ratio

\[
Q\equiv\frac{f_s}{f_s^{(\rm ideal)}}=\frac{dE^{(\omega)}/d\omega^2
|_{\omega=0}}{dE^{(\omega)}_{\rm ideal}/d\omega^2|_{\omega=0}}=\left.
\frac{dE^{(\omega)}}{d\omega^2}\right|_{\omega=0}
\frac{M^2\omega_0^2}{4\pi^2JN}.
\]

\noindent If now one takes into account the relation between the
cranking angular velocity and the `twist' $\Theta$ used
in~(\cite{Keith1}), namely $\Theta=2\pi\omega/\omega_0$, to express
the quenching factor $Q$ in terms of $dE^{(\omega)}/d\Theta^2$, one
obtains the expression used there for the superfluid fraction.

\section{Properties related to condensation}\label{sec:condensation}

\begin{figure}
    \includegraphics[scale=0.4]{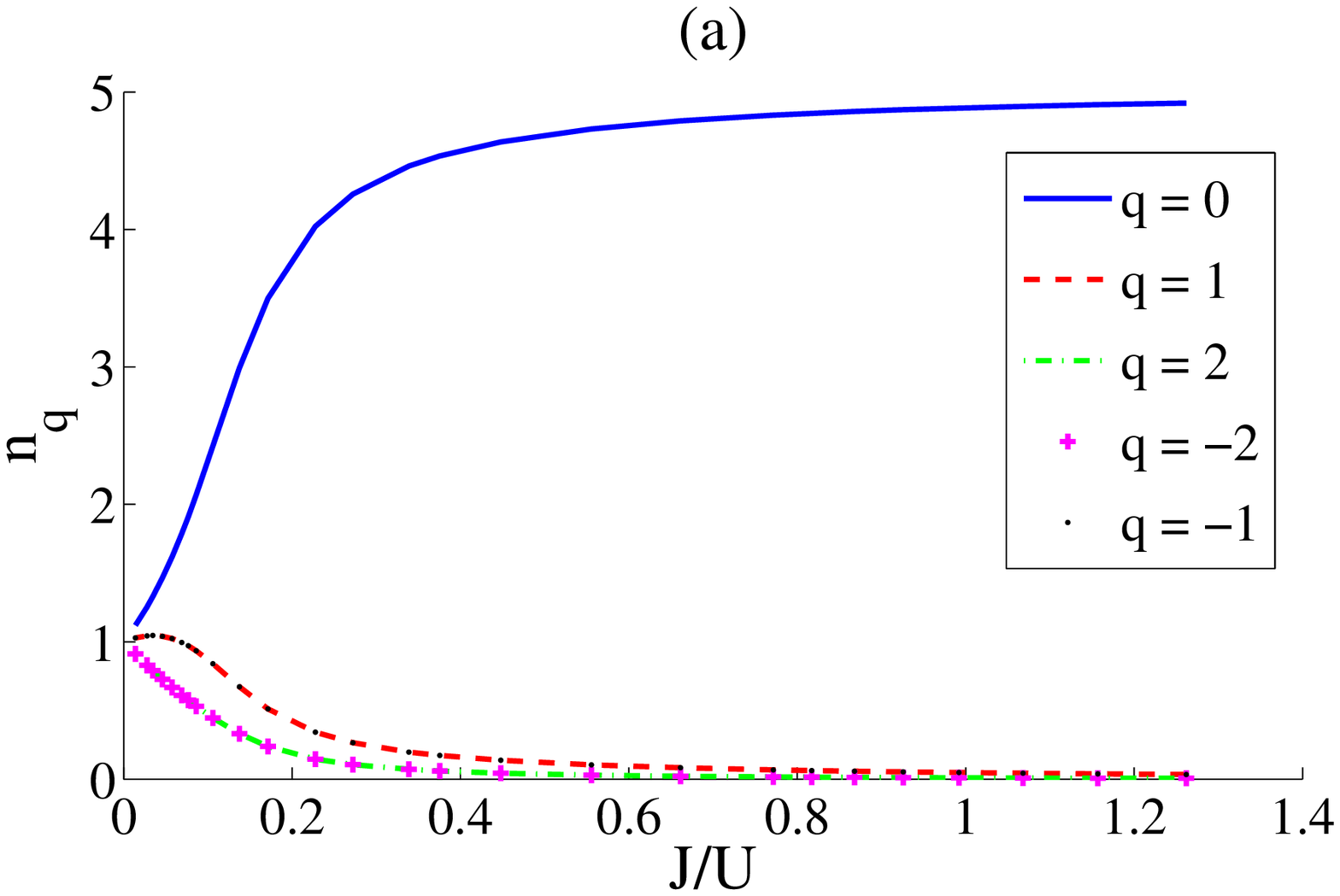}
    \includegraphics[scale=0.4]{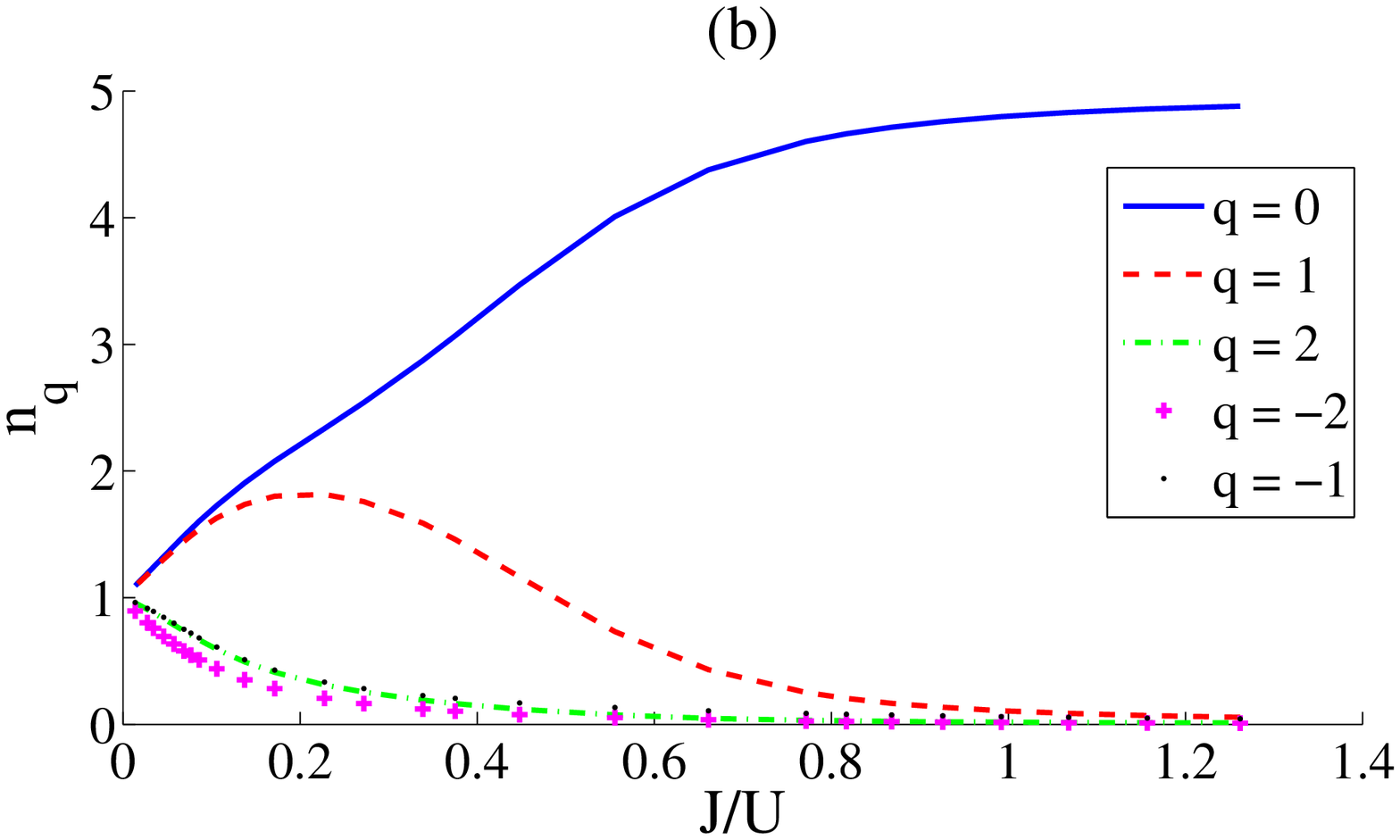}
    \includegraphics[scale=0.4]{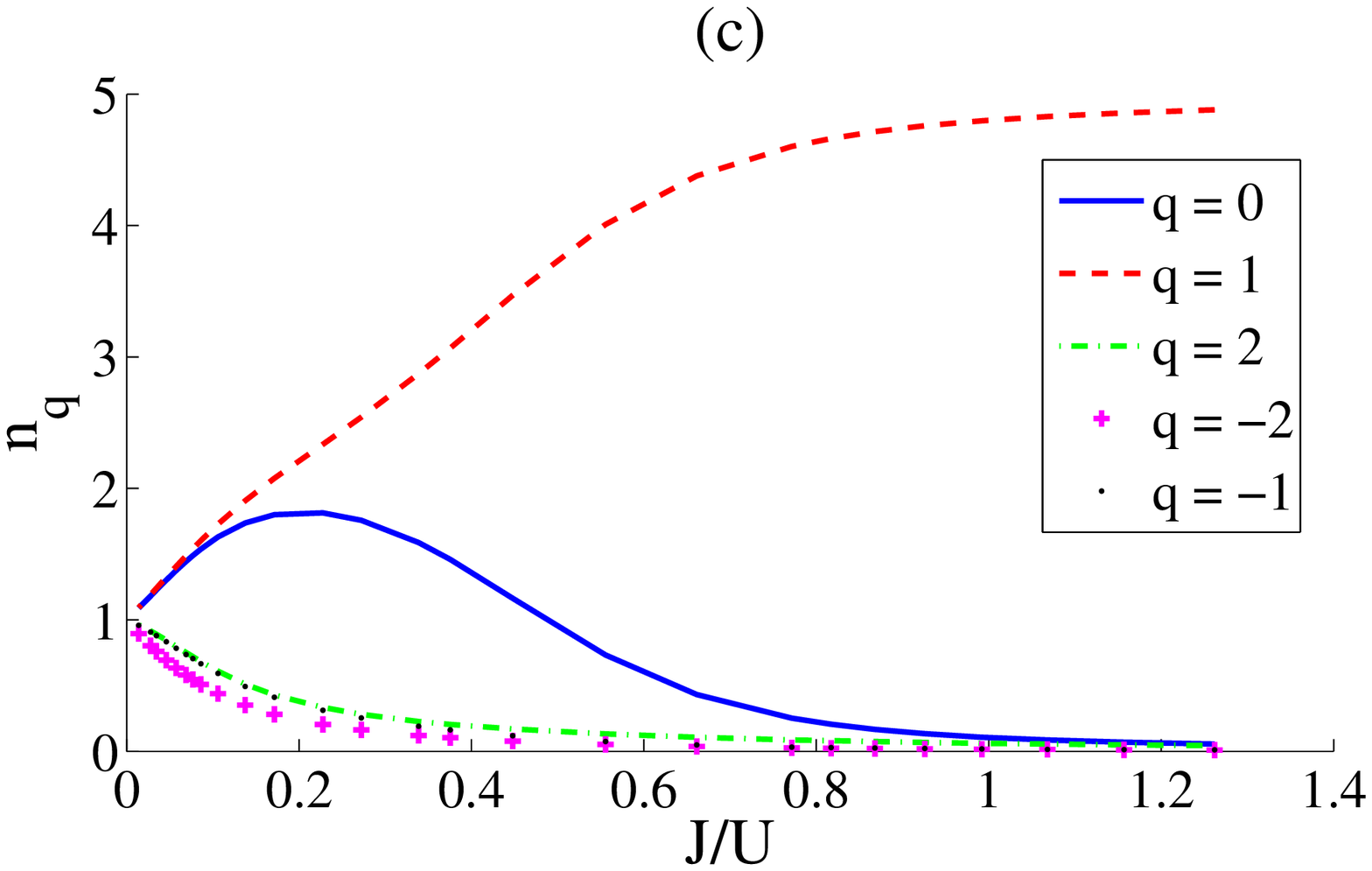}
    \includegraphics[scale=0.4]{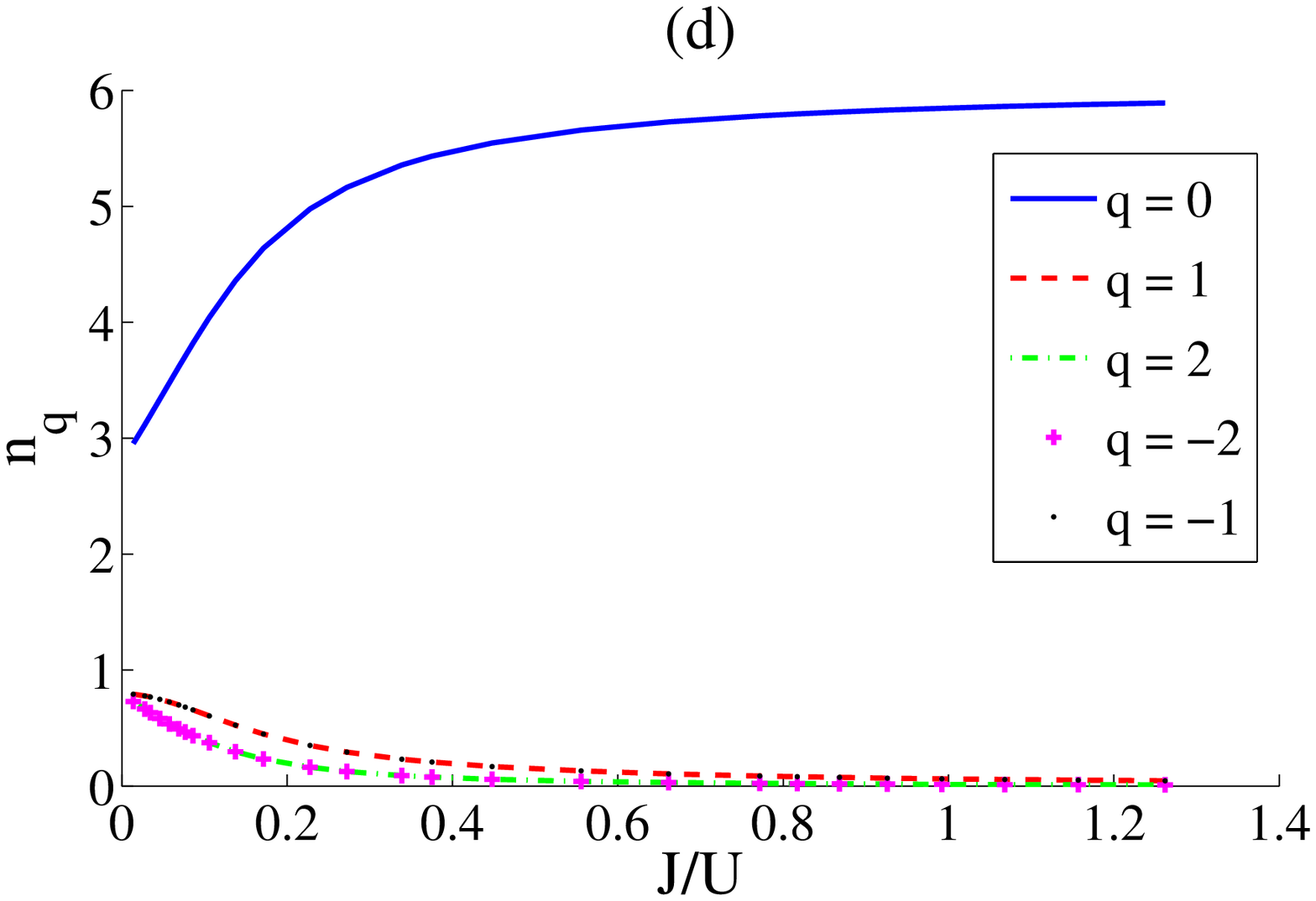}
    \caption{(Color online) Eigenvalues of the reduced density matrix
      for $\Lambda = 0.06\epsilon_0$ and $M = 5$. In (a), (b) and (c)
      the mean filling per site $N/M = 1$, while in (d), $N/M = 1.2$
      . In (a), $\omega = 0$ while in (b), $\omega = 0.48\omega_0$ and
      in (c), $\omega = 0.52\omega_0$, i.e.  just before and after the
      single-particle level crossing at $\omega = 0.5\omega_0$. At
      $\omega=0.5\omega_0$ there is a migration of population from the
      state with quasi-momentum $q = 0$ in (a) and (b) to the one with
      $q = 1$ in (c); (d) illustrates qualitative differences in the
      limit $J/U \rightarrow 0$ between the incommensurate and the
      commensurate situations.}
\label{fig:occ}
\end{figure}
Eq.~(\ref{Ifromj}) and its interpretation that full coherence of the
superfluid current is restored only in the limit where only one of the
natural orbitals is occupied suggests that further attention should be
given to the eigenvalues and eigenvectors of the one-body density
matrix. In fact, their properties are related to the Penrose and
Onsager criteria~\cite{Onsager} for characterizing
Bose-Einstein condensation in systems of interacting bosons. It is 
defined in terms of the macroscopic occupation of a natural orbital,
which then plays the role of `condensate wavefunction' (see
Eq.~(\ref{b1}) and discussion on the spectral decomposition of the
reduced density matrix in Sec.~(\ref{sec:superfluidity})). Barring
important limiting procedures which have to be used when dealing with
extended systems, and notwithstanding non-existence theorems for
Bose-Einstein condensation as a phase transition in one dimensional
systems~\cite{Giamarchi}, in the present case of fixed, finite number
of bosons the behavior of the natural orbital occupation numbers with
the parameters of the model Hamiltonian is still revealing of
condensation oriented one-body properties of the correlated many-boson
state.

Typical behavior for occupation numbers of the system's ground state
natural orbitals is shown in the plots of Fig.~\ref{fig:occ} (a), (b)
and (c) for a situation with commensurate filling (the number of
bosons $N$ being an integer multiple of the number of sites in the
array, $M$), and in Fig.~\ref{fig:occ} (d) for an incommensurate
case. In all cases the two-body interaction parameter has been fixed
at $\Lambda=0.06\epsilon_0$. Use of the larger value
$\Lambda=0.3\epsilon_0$, as in Figs. \ref{fig:current} and
\ref{fig:suplobes}, gives however essentially identical results,
reflecting the validity of the tight binding approximation in this
context. Since in the incommensurate case one may have to deal with
different values of total quasi-momentum due to the
$\omega$-dependence of the Bloch energies, the two cases are better
discussed separately.

We start with the commensurate situation where, regardless of the
values of $\omega$, the ground state of the many-body system is always
found in the subspace of total quasi-momentum $Q_T = 0$. As shown in
Figs.~\ref{fig:occ} (a), (b) and (c), evaluated for the case $N=M$, in
the extreme tight-binding limit $J \rightarrow 0$, one has a fully
degenerate situation in which each of the orbitals has unit
occupation, implying that the reduced one-body density is an {\it
  incoherent} superposition with equal weights of one body densities
constructed from each of the first band Bloch functions. This is in
fact a consequence of the localization of each particle in one of the
sites, achieved through probability amplitudes involving the (Wannier)
{\it coherent} superposition of the Bloch functions. In the
commensurate case with $\rho_0=1$, localization causes the two-body
energy to vanish in the limit $J\rightarrow 0$. Increasing the
effective hopping parameter $J$ by lowering the barrier strength leads
eventually to a situation in which the trace of the reduced one-body
density is carried by essentially a single eigenvalue, the associated
natural orbital being the Bloch state with lowest single-particle
energy at the considered value of the cranking angular velocity
$\omega$. This implies that the spectral decomposition of the reduced
one-body density essentially reduces to a single term, which
corresponds to the underlying many-body state approaching the simple
form of a product state in which all bosons are coherently delocalized
in the collectively occupied Bloch wavefunction. Parts (b) and (c) of
Fig.~\ref{fig:occ} illustrate the change of role of Bloch orbitals
near the level crossings at $\omega=0.5\omega_0$.

Fig.~\ref{fig:occ} (d) shows the evolution under increasing effective
hopping parameter $J$ of the eigenvalues of the reduced one-body
density in the incommensurate case $N=6$, $M=5$ for $\omega=0$. Unlike
in the commensurate case, here one no longer has occupation number
degeneracy for the natural orbitals at the extreme tight-binding limit
$J\rightarrow 0$, but an occupation enhancement of the lowest Bloch
state which in fact typically exceeds the contribution of the
extranumerary boson, an effect which may be traced to the action of
bosonic enhancement factors~\cite{PizaDiagrams}. The delocalization
process induced by hopping, associated with progressive dominance of a
single Bloch state, is however maintained.

Finally, it is worth noting that, in all cases, the scale over which
the transition to delocalization occurs, in terms of $J/U$ coincides
with that which may be associated with the extension of the Mott
insulator lobes in the ground-state phase diagram (see
Fig.~\ref{fig:mottlobes}), suggesting the connection of these two
features. 

\subsection{Discussion}

The relatively small values obtained for the superfluid fraction
throughout the range of barrier strengths in which the delocalization
transition takes place is perhaps not surprising if one keeps in mind
that this range falls within the tight-binding domain in which the
Bose-Hubbard model is an excellent approximation to the present
treatment of the Hamiltonian~(\ref{5}), and that the potential
barriers themselves constitute an important mechanism coupling the
externally imposed rotation to the dynamics of the many body
system. From this point of view the obtained values for the superfluid
fraction based on the two-fluid model arise from a current, in the
rotating frame of reference in which the lattice potential is at rest,
that still manages to flow through the hindrance of the potential
barriers.

There are two features of this current that deserve some
remarks. First, the fact that in the present one-dimensional circular
geometry it is independent of position on the circle (see
Eq.~(\ref{sol})) is clearly a necessary consequence of stationarity,
which in particular requires a time independent one-body
density. Moreover, the restriction to the Bloch orbitals of the first
band, together with the (modular) conservation of quasimomentum by the
Hamiltonian~(\ref{5}), imply the vanishing of mean field effects of
the two-body force, thus causing the first band Bloch states
themselves to play the role of eigenstates of the reduced one-body
density. From the one-body equation leading to them it follows that
each of the individual square brackets in Eq.~(\ref{sol}) vanishes for
the numerical calculations reported here. It should be kept in mind,
however, that this is just an artifact of the adopted truncation of
the single particle basis, and not true in the general case in which,
according to Eq.~(\ref{sol}), only the appropriately weighted sum of
them is independent of $\varphi$. 

The second feature deserving comment is the fact that the current
associated with the superfluid fraction is generally given as an
incoherent sum of contributions associated with each of the populated
eigenstates of the reduced one-body density. Thus, to the extent that
it favors the occupation of a single such eigenstate, the
delocalization transition associated with the closing of the Mott
insulator lobes effectively promotes the complete coherence of the
full current. In the context of the calculations reported here, the
effects of incoherence stem from the different values of the
($\varphi$-independent) contributions of the first band Bloch states
together with the respective occupation numbers. The special role
played by the Bloch states is however also an artifact of the adopted
truncation of the single-particle base. In general, further many-body
dynamical effect will be encoded in the structure of the eigenstates
of one-body density.

\section{Conclusions}\label{sec:conc}
We presented a numerical study of properties of a small system of
identical bosons in one-dimensional circular lattice undergoing
rotation. The calculations were based on the diagonalization of the
standard many-body Hamiltonian in the external lattice potential with
contact two-body effective interactions truncating the involved single
particle basis to the first band Bloch states of the rotating lattice.
With the appropriate scaling of the system parameters, results agree
quantitatively with those obtained from the simple Bose-Hubbard model
in the tight-binding limit, but will elsewhere incorporate additional
effects (such as modified single-boson energy spacings and hopping
effects other than nearest neighbors), subject to the limitations
imposed by the adopted truncation (to the first band) of the
single-particle orbitals. 

Results obtained for a reinterpretation of the (grand-canonical)
ground state phase diagram in terms of `separation energies' obtained
by comparing ground state energies of systems with different numbers
of particles reproduce results of known grand-canonical calculations
quantitatively even for a small lattice, including the `reentrant
behavior' of the Mott insulator lobes characteristic of
one-dimensional systems. The closing of the lobes is clearly
associated with the delocalization of the bosons over the lattice which
results from the progressive occupation of a single eigenstate of the
one-body reduced density matrix.

Numerical results for ground state superfluid fractions, obtained in
the context of two-fluid phenomenology, link the closing of the Mott
insulator lobes with increased coherence of a uniform, solenoidal
single particle current associated with the intrinsic inertia of the
superfluid fraction. This association can be formally extended to
three dimensional traps, also independently of assuming the existence
of a condensed mode or any particular currents associated with it.

The value of the superfluid fraction remains small throughout
the parameter domain associated with the transition, and grows
monotonically saturating at the value 1 only with the complete
smoothing out of the lattice. This suggests that the `Mott-insulator
to superfluid transition' may be a transition related rather to
properties of coherence and delocalization, than to superfluidity in
terms of propensities in the system to physical flow.

FP thanks Jonas Larson and Jani-Petri Martikainen for useful comments
and discussions. Work has been supported by FAPESP and the Swedish
Research Council (Vetenskapenr\aa det).

\end{document}